\documentclass{iau}
\usepackage{graphicx} 
\usepackage{comment}

\title[IAUS291.~~LOFAR LBA single-pulse study of the PSR~B0809+74] %% short title %%
{Pulsar emission at the bottom end of the electromagnetic spectrum} %% full title %%

\author[V. Kondratiev et al.]  %% short author list %%
{Vladislav Kondratiev$^1$ and the LOFAR Pulsar Working Group}

\affiliation{$^1$ASTRON, the Netherlands Institute for Radio  Astronomy, \\ 
Postbus 2, 7990 AA Dwingeloo, The Netherlands \\ email: {\tt vlad.kondratiev@gmail.com}}

\pubyear{2012}
\volume{291}  %% insert here IAU Symposium No.
\jname{\mbox{Neutron Stars and Pulsars: Challenges and Opportunities after 80 years}}
\editors{J. van Leeuwen, ed.} 
\begin{document}

\maketitle

%% -- Abstract ----------------------------------
\begin{abstract}
Pulsars are arguably the only astrophysical sources whose emission spans the entire electromagnetic
spectrum, from decameter radio wavelengths to TeV energies.  The LOw Frequency ARray (LOFAR) offers the unique possibility 
to study pulsars over a huge fractional bandwidth in the bottom 4 octaves of the radio window, from 15--240~MHz. 
Here we present a \mbox{LOFAR} study of pulsar single pulses, focussing specifically on the bright nearby pulsar B0809+74. 
We show that the spectral width of bright low-frequency pulses can be as narrow as 1~MHz and scales with increasing 
frequency as $\Delta f/f_\mathrm{c} \sim 0.15$, at least in the case of the PSR~B0809+74. This appears to be intrinsic 
to the pulsar, as opposed to being due to propagation effects. If so, this behavior is consistent with predictions by
the strong plasma turbulence model of pulsar radio emission. We also present other observed properties of the single 
pulses and discuss their relation to other single-pulse phenomena like giant pulses.
%% add here a maximum of 10 keywords, to be taken form the file <Keywords.txt>

\end{abstract}

% add below any authors, subjects and objects for indexing 
%   add more lines if necessary
%   but leave all lines commented out
%\index[author]{LastName1, Initials|textbf}
%\index[author]{LastName2, Initials|textbf}
%\index[subject]{Keyword1}
%\index[subject]{Keyword2}
%\index[object]{Object1}
%\index[object]{Object2}

\firstsection % if your document starts with a section,
              % remove some space above using this command.
\section{Introduction}

In recent years a number of studies of strong individual pulses from
bright nearby pulsars were performed, including some at low radio
frequencies (see, e.g., \cite[Kuzmin 2006 and references
  therein]{k06}; \cite[Weltevrede et al. 2006]{w+06};
\cite[Karuppusamy et al. 2011]{k+11}).  Higher radio frequencies are
more favorable for pulsar studies because of several factors, such as
intra-channel dispersion, scattering, the Galactic synchrotron
background, and the ionosphere. Nonetheless, the very low-frequency
range is interesting for pulsar studies because of pulse-profile
evolution, which can become quite dramatic towards the lowest
frequencies (\cite[Hassall et al. 2012]{hsh+12}). Next, 
some phenomena are potentially not seen at higher frequencies (or 
become less  pronounced there), in particular the anomalously
bright individual pulses from five nearby pulsars including
PSR~B0809+74, reported by \cite[Ulyanov et
  al. (2006)]{u+06} 
based on observations with the UTR-2 radio telescope. It is not clear whether these pulses
are attributed to giant pulses or other single-pulse phenomena
observed from other pulsars at higher frequencies. In these
proceedings we report the first broadband low-frequency single-pulse
study of the PSR~B0809+74 using the LOFAR telescope.

\section{Observations and Data processing}

Observations were carried out using the Low Band Antennas (LBAs) in
the LBA\_OUTER array mode, where the 48 outermost dipoles in each LBA
field are used. The instantaneous bandwidth of 46.875~MHz centered at
a frequency of about 38~MHz was split into 240 subbands of 32
channels each and originally sampled every
$491.52~\mu$s.
\begin{figure}[tbhp]
\begin{center}
\includegraphics[height=\textwidth,angle=270]{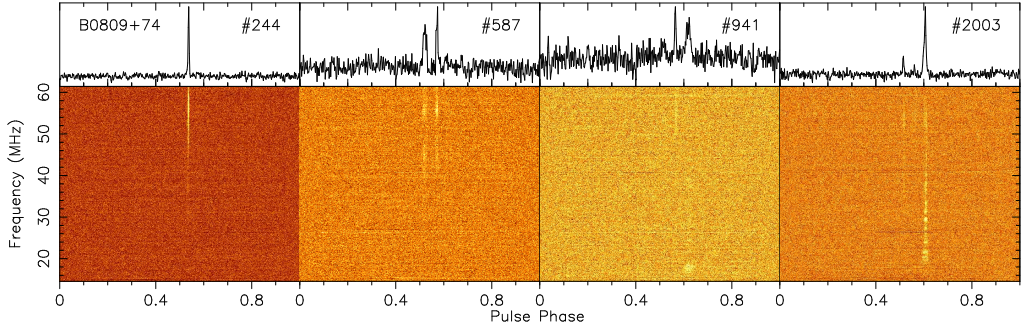}
\caption{Examples of characteristic spectra and profiles of individual pulses of the pulsars B0809+74. 
The number in the top-right of each panel shows the pulse number from the beginning 
of the observation. There are 438 bins in the period, and each spectrum is comprised of 480 channels of 97.656~kHz each.} 
\label{0809-sample}
\end{center}
\end{figure}
Each observation was the coherent sum of the six ``Superterp''
stations, CS002--CS007. A detailed description of LOFAR's pulsar observing modes and the online data reduction 
pipeline is given in \cite[Stappers et al. (2011)]{sha+11}.

The data were converted from the LOFAR beam-formed format to the {\tt
  PSRFITS} data format, corrected for the varying
gain over the LBA band,  and processed with {\tt DSPSR}\footnote{\tt
  http://dspsr.sourceforge.net} to form both 
an integrated profile as well as single-pulse integrations. We used
{\tt pdmp} from {\tt PSRCHIVE}\footnote{\tt
  http://psrchive.sourceforge.net} on a sample of strong pulses to
determine the best DM for the time of our observations: 5.752~pc cm${}^{-3}$, with the DM jitter between the pulses
being on the order of 0.002~pc cm${}^{-3}$.
Many of the pulses showed very narrow-band structure in their spectra
(see Figure~\ref{0809-sample}, and details below). 
Using a 2D time-frequency search technique (Kondratiev et al., in prep)
we next analysed the spectra of the complete
sample of single pulses.

\section{Results}

The complete results of our study will be presented in Kondratiev et
al.\ (in prep); here we briefly present a few highlights.

Figure~\ref{0809-sample} shows some typical pulses from the
PSR~B0809+74. In a given pulse, emission can occur in either the
leading or trailing component; in both components at the same time; or
sometimes even in three components corresponding to three
subpulse-drift bands (see, e.g., pulse $\#941$ on
Figure~\ref{0809-sample}). The frequency structure of the pulse spectra
can be also quite different. For instance, pulse $\#941$ shows a very
narrow emission patch at 19~MHz in the second component, but the
strongest emission from other two components occurs at the higher
frequencies.  Other examples (\eg pulse $\#2003$) show very broadband
structure comprised of a number of smaller, individual patches with the
maximum tending to be at lower frequencies ($<30$~MHz). On average the
pulse spectra of the PSR~B0809+74 are quite narrow and constitute only
half of the band.

\begin{figure}[tbhp]
\begin{center}
%\floatbox[{\capbeside\thisfloatsetup{capbesideposition={right,top},capbesidewidth=3.5cm}}]{figure}[\FBwidth]
\includegraphics[clip,trim=9mm 4mm 18mm 14mm,
  width=0.7\textwidth]{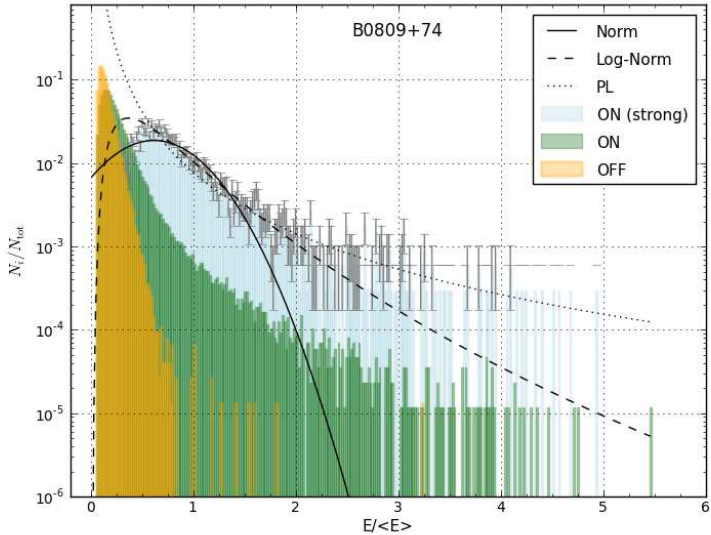}
\caption{Pulse energy distribution for B0809+74. 
Energies are for the emission patches over patch's frequency width normalized by the energy of the average
profile over the entire band. Distributions in green and yellow are for ON-pulse and OFF-pulse phase windows,
respectively, using the same search technique with the threshold of $1\sigma$. The distributions in light blue
are for the strong patches with the spectral flux density over $5\sigma$.}
\label{energy_distrib}
\end{center}
\end{figure}

Figure~\ref{energy_distrib} presents the pulse energy distribution for
the pulsar B0809+74.  The energy is calculated for the emission
patches of every pulse component in the profile rather than in the
entire band due to the presence of narrow-band pulses. Distributions
are skewed towards higher energies due to the search technique as we
searched for {\it highest} spectral flux density. It is clear that
distributions are very similar at low energies with slightly less
number of low-energy patches in the ON-pulse window. A few positive
outliers for the OFF-pulse histogram indicates the presence of RFI but
their number is insignificant. We also show the ON-pulse histogram for
the strong patches with the spectral peak flux density $>5\sigma$. The
roll-off of the histogram at low energies represents the selection
bias for patches with peak spectral fluxes close to the detection
threshold.  The apparent larger fraction of stronger patches than
$1\sigma$-patches for the same energy, is caused by
the fact that overall number of found patches is much larger for the
$1\sigma$ than the $5\sigma$ threshold.

We fit a normal, log-normal and power-law distribution with low-energy cutoff. For our fits we excluded 
the very high energy tail of the distributions that have small statistics. 
The lognormal distribution provides the best goodness of fit. However, fitting the power-law distribution 
to the high-energy tail for energies $>2.4\langle E\rangle$
provides better results. In general, the significance of the fit is not high enough for both pulsars to reject neither lognormal
nor power-law distributions, and the better statistics is needed.

\subsection*{The narrow-band emission of the PSR~B0809+74}

The narrow-band spectra of some of the pulses from the PSR~B0809+74
are quite unique and have never been observed  at higher frequencies. We
found it very unlikely that this narrow-band frequency pattern is
caused by either ISM scintillations or ionosphere. For the former
estimates of the decorrelation bandwidth give values of $\lesssim
2$~kHz, so all scintillations should be completely averaged out in our
data. For the ionosphere the rate of change in its state is long
(mins--hours) compared to time scales of about a second for the use
case when narrow-band pulses follow each other.

\begin{figure}
 \begin{minipage}[b]{0.66\textwidth}%
 \centering
 \includegraphics[clip,trim=9mm 4mm 15mm 13mm, width=\textwidth]{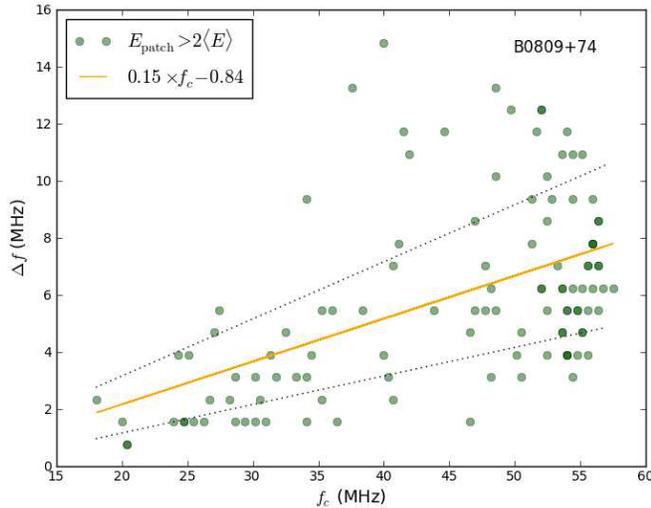}
 \vspace{-10mm}%
\end{minipage}%
\begin{minipage}[b]{0.33\textwidth}%
 \caption{Dependence of the patch spectral width on their emission frequency (central frequency of the
patch) for the pulsar B0809+74 for the all strong emission patches with the energy $E>2\langle E\rangle$ (green circles). 
The orange line, $\Delta f/f_\mathrm{c}\sim0.15$, shows the best least-square fit. The dotted lines show the 
same dependency but for the slope of 0.1 and 0.2.}
\label{0809df-f}
\end{minipage}%
\vspace{4mm}%
\end{figure}

Figure~\ref{0809df-f} shows the dependence of the patch's frequency
width on their central frequency for the pulsar B0809+74 for the
strong patches with the energy $E>2\langle E\rangle$. Though data
points are somewhat scattered, they qualitatively lie on the line,
$\Delta f/f_c \sim 0.15$. This agrees very well with the prediction
from the strong plasma turbulence (SPT) model (\cite[Weatherall
  1998]{w98}).  The SPT model of pulsar radio emission predicts
narrow-band radiation with $\Delta f/f \sim 0.1-0.2$.  If indeed true,
this would also provide a direct link between giant pulses and
anomalously intensive pulses observed at low frequencies.

\section{Summary}

We observe occasional bright narrow-band pulses from the PSR B0809+74 at frequencies 15--62~MHz. 
Their spectra can be as narrow as 1~MHz and tend to have a width of 3~MHz.

We identified pulse sequences from the pulsar B0809+74 where narrow patch of emission is drifting up in 
frequency from pulse to pulse. We see evidence for similar frequency drift for at least one other pulsar.

The origin of these narrow-band pulses is likely to be pulsar-intrinsic rather than due to propagation 
effects in ISM or ionosphere. 

At the moment, the observed pulse properties of the PSR~B0809+74 do not allow us to relate low-frequency 
bright narrow-band pulses to ``spiky emission'', or to either giant pulses or regular emission.

For PSR~B0809+74, the spectral width of the strong emission patches
scales with increasing frequency  as $\Delta f/f_\mathrm{c}\sim 0.15$,
qualitatively agreeing with the prediction of the SPT model. 
This supports the relation between bright narrow-band pulses at low frequencies and giant pulses. 

~\\
\noindent{\bf Acknowledgements}\\
LOFAR, the Low Frequency Array designed and constructed
by ASTRON, has facilities in several countries 
that are owned by various parties (each with their own
funding sources), and that are collectively operated by
the International LOFAR Telescope (ILT) foundation under
a joint scientific policy.

\end{document}